\documentclass[aps,prb,floats,floatfix,twocolumn]{revtex4-1}

\usepackage{graphicx}
\usepackage{epstopdf}
\graphicspath{{./Figs/}}

\usepackage{latexsym}
\usepackage{amsmath}
\usepackage{amssymb}
\usepackage{bm}
\usepackage{wasysym}

\usepackage{ifthen}
\usepackage{color}
\usepackage[colorlinks,allcolors=blue,bookmarks=true]{hyperref}
\usepackage[percent]{overpic}

\newcommand{\eexp}[1]{\mathrm{e}^{#1}}

\newcommand{\braket}[1]{ \left\langle #1 \right\rangle}

\newcommand{\be}[1]{\begin{eqnarray}{\label{e#1}}} 
\newcommand{\beq}{\begin{eqnarray}}
\newcommand{\eeq}{\end{eqnarray}} 

\newcommand{\hide}[1]{}
\newcommand{\Eq}[1]{\textcolor{blue}{{Eq.}\!\!~(\ref{#1})}} 
\newcommand{\App}[1]{\textcolor{blue}{{Appendix}\!~#1}} 

\newcommand{\Fig}[1]{\textcolor{blue}{Fig.}\!\!~\ref{#1}}

\newcommand{\hrefl}[2]{\href{#2}{(#1)}}
\newcommand{\paragraphtitle}[1]{{\bf #1.--}}

\begin{document}

\title{Superfluidity in Bose-Hubbard circuits}

\author{Geva Arwas, Doron Cohen}

\affiliation{Department of Physics, Ben-Gurion University of the Negev, P.O.B. 653, Beer-Sheva 84105, Israel}

\begin{abstract}
A semiclassical theory is provided for the metastability regime-diagram
of atomtronic superfluid circuits. Such circuits typically
exhibit high-dimensional chaos; and non-linear resonances that couple
the Bogoliubov excitations manifest themselves. Contrary to the
expectation these resonances do not originate from the familiar Beliaev
and Landau damping terms. Rather, they are described by a variant of the
Cherry Hamiltonian of celestial mechanics. Consequently we study the
induced decay process, and its dependence on the number 
of sites and of condensed particles.
\end{abstract}

\hide{
Metastability is a major theme in physics. Thermodynamically 
it is associated with the existence of local minima 
in the energy landscape of the system under consideration.
A very different notion of ``dynamical stability" applies with 
regard to the very ``big" dissipationless celestial mechanics. 
But in-fact it applies also with regard to the very ``small" of mesoscopic physics. 
Here our interest is in the most prominent example in condensed-matter physics, 
the metastability that is known as ``superfluidity".
Atomtronic circuits, that are described by the Bose-Hubbard Hamiltonian, 
are regarded as a paradigm for quantized high-dimensional chaos.
We show that nonlinear resonances, that couple the Bogoliubov excitations,
manifest themselves. Contrary to the expectation, these resonances do not
originate from the familiar Beliaev and Landau damping terms. Rather,
they are described by a variant of the Cherry Hamiltonian of celestial mechanics, 
and can lead to hyperbolic decay of the superflow. 
}

\maketitle


\section{Introduction}

Metastability is a major theme in physics. The first picture that comes
into mind is that of local minima in the energy landscape of the system
under consideration. Possibly this point of view is correct in a
thermodynamic context, where dissipation always drives the system down
into valleys. But there is a different notion of ``dynamical stability"
that applies with regard to the very ``big", celestial mechanics, where
dissipation is negligible. Surprisingly it is not yet widely recognized
that it applies also with regard to the very ``small", where the
technological challenge is again to minimize the effect of dissipation.

Our interest is in the most prominent example in condensed-matter
physics, namely, the metastability which is known as ``superfluidity".
Recent experimental studies have focused on low dimensional superfluid circuits \cite{NIST2,hadzibabic,BoshirPRL2013,baker} in a toroidal geometry.
A challenge is to realize a discrete ring version \cite{Amico2014} 
that is hopefully described by the $M$-site Bose-Hubbard Hamiltonian (BHH). 
This will allow the design of SQUID-like qubits or related devices \cite{anamaria1,anamaria2,brand2,brand1,Amico2014,Aghamalyan15}.

The hallmark of superfluidity is the possibility to witness a {\em metastable}
persistent current, aka flow state. 
This notion of superfluidity does not assume a thermodynamic limit, 
and is well-defined even in the absence of a phase transition.
We focus the analysis below on one-dimensional BHH circuits 
with a {\em finite} number of sites. 

The stability of a superflow is a widely studied theme.
In a uniform potential the Landau criterion \cite{landau,critVel1} leads to a critical velocity, 
below which the superfluid is robust against perturbations. 
The system is then said to be ``Landau stable" or ``Energetically stable". 
Later it was shown theoretically \cite{smerzi,niu,polkovnikov2005decay} and confirmed experimentally \cite{dynStab1,dynStab2,dynStab3} 
that under a periodic potential, as in optical lattices,  
the superflow can survive due to ``dynamical stability". 
The critical velocity there is determined by calculating the Bogoliubov excitation spectrum, 
which is in essence linear-stability analysis.

It has been established in our recent work \cite{sfc} that one has to go beyond 
the linear-stability reasoning in order to analyze the metastability of currents in rings that 
have ${M=3}$ sites, which is like dealing with ${d=2}$ degree of freedom dynamical system. 
The major ingredient that comes into play is ``chaos".  
The BHH is formally a quantized chaotic system, see e.g. \cite{kolovskiPRL,kolovskiPRE,KolovskyBlochOsc},
and therefore the metastability of the superflow is not merely a matter 
of linear-stability analysis. 
It is the purpose of the present work to explore the physics 
of superfluidity for ${M>3}$ rings, 
where the underlying phase-space features {\em high dimensional chaos}
with ${d>2}$ degrees of freedom, and intricate web of non-linear resonances.

The idea that a linearly-stable system can be in-fact dynamically unstable, 
due to non-linear resonances, is almost a century old \cite{LLbook,chirikov1960}, 
but has attracted very little attention in the condensed matter literature, 
with few exceptions \cite{NonlinInstPRL2015}.
One of the first examples for that has been proposed by Cherry \cite{Cherry}. 
The Cherry Hamiltonian appears in celestial mechanics, 
in the context of the restricted 3-body problem \cite{meyer2008introduction}, 
and also in plasma physics, in the context of negative energy modes \cite{Pfirsch1993,morrison1,morrison2}.

In the present work we would like to study how superfluidity of an atomtronic circuit 
is affected by non-linear resonances, in regimes where the dynamics is traditionally 
considered to be stable via linear analysis. Our work is oriented towards
hysteresis-type experiments similar to that of \cite{NIST2}. 
We would like to provide for such future experiments superfluidity regime diagrams, 
and to clarify how phase-space features are reflected in the time dependence 
of the decay process following a quench scenario. 
In a broader perspective we would like to demonstrate that tools 
of analytical mechanics and semiclassics are extremely advantageous 
in an arena that is largely dominated by formal field-theoretical many-body methods.

\ \\
 
\paragraphtitle{Outline:}
We introduce the BHH in Section~II; 
provide the semiclassical perspective on metastability in Section~III, 
and clarify how it can be probed by a quench experiment in Section~IV.
Then we explore how stability is engendered by nonlinear resonances in section~V,  
while on the other hand we explain the emergence of a stability island off-resonance in Section~VI. 
Finally we expand on the analysis of the decay process in Section~VII,  
which leads naturally to the categorization of the various 
types of metastability in the present context. 
Several Appendices provide extra technical details.

\clearpage

\begin{figure*}
\begin{center}
\begin{overpic}[width=0.32\hsize,tics=10]{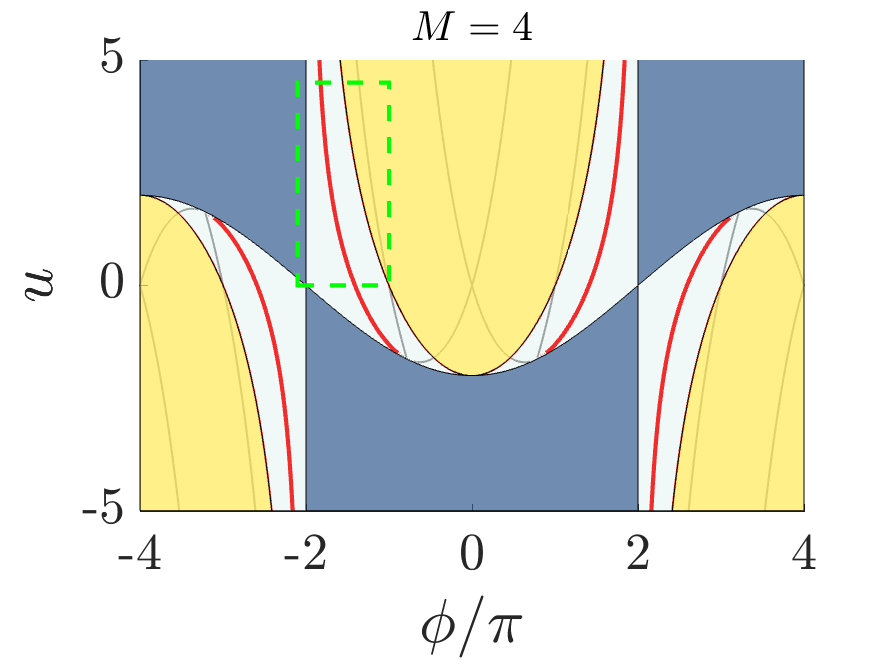}
 \put (10,75) {(a)}
\end{overpic}
\begin{overpic}[width=0.32\hsize,tics=10]{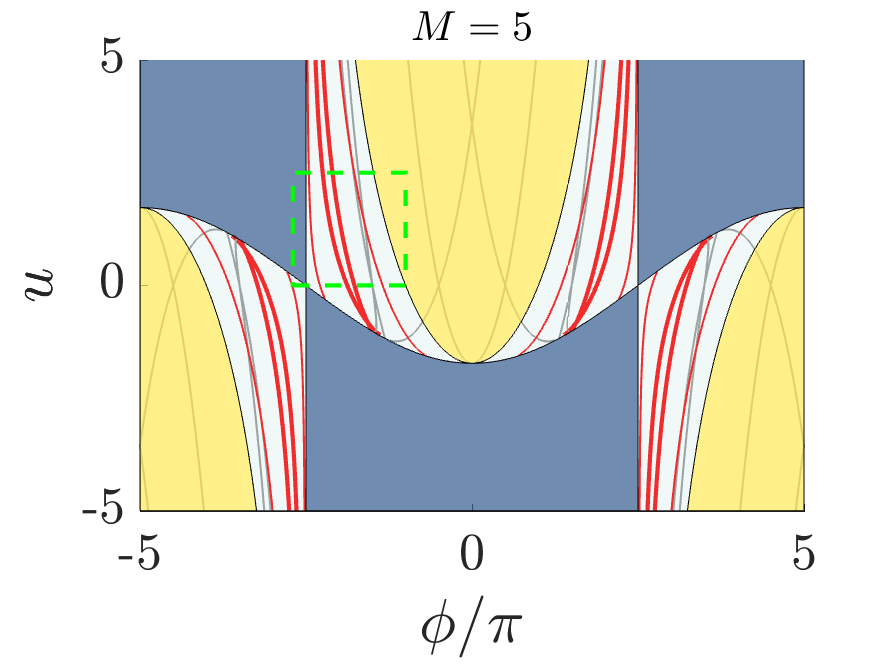}
 \put (10,75) {(b)}
\end{overpic}
\begin{overpic}[width=0.32\hsize,tics=10]{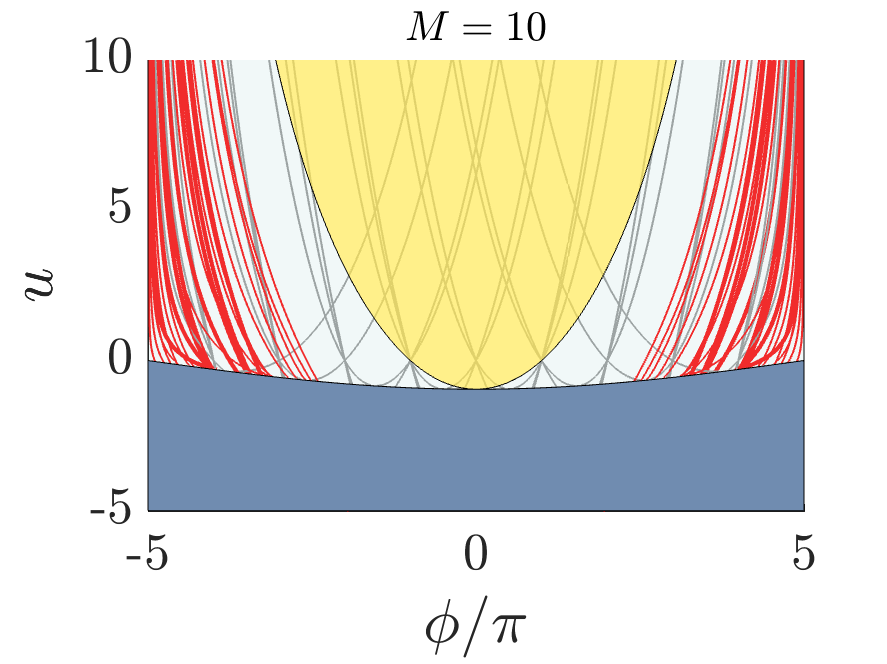}
 \put (10,75) {(c)}
\end{overpic}
\caption{ \label{fg1} 
{\bf The superfluidity regime diagram}.
The panels are for rings with $M=4,5,10$ sites. 
Yellow and gray colors indicate the Landau-stability and linear-instability regions.
Non linear resonances of the $V^A$ and $V^B$ type are indicted by red and gray lines respectively. 
The explicit expression for the ``red" resonance in panel (a) is \Eq{e9}.
The additional thin red lines in panels (b,c) are 4th order resonances. 
The green squares mark regions of interest that will be explored in \Fig{fg2}. 
For sake of generality we consider both positive and negative values of~$u$.
} 
\end{center}
\end{figure*}

\section{Circuit Hamiltonian}

The BHH is a prototype model of cold atoms in optical lattices 
that has inspired state-of-the-art experiments 
with condensed particles \cite{BHH1,BHH2}. 
For $M$ sites, in a ring geometry:
\be{1} 
\mathcal{H} = \sum_{j=1}^{M} \left[
\frac{U}{2} \bm{a}_{j}^{\dag} \bm{a}_{j}^{\dag} \bm{a}_{j} \bm{a}_{j} 
- \frac{K}{2} \left(\eexp{i(\Phi/M)} \bm{a}_{j{+}1}^{\dag} \bm{a}_{j} + \text{h.c.} \right)
\right] \ \ \ \ \ 
\eeq
where $K$ is the hopping frequency, $U$ is the on-site interaction, 
and $j$ mod$(M)$ labels the sites of the ring. 
The $\bm{a}_{j}$ and $\bm{a}_{j}^{\dag}$ are the Bosonic annihilation and creation operators, 
and the $\bm{n}_j \equiv \bm{a}_{j}^{\dag}\bm{a}_{j}$ are occupation operators. 
The total number of particles $N=\sum \bm{n}_j$ is a constant of motion.
The so-called Sagnac phase $\Phi$ is proportional to 
the rotation frequency of the device: it can be 
regarded as the Aharonov-Bohm flux that is associated
with Coriolis field in the rotating frame~\cite{fetter,NIST1}.
Optionally it can be realized as a geometric phase using artificial gauge fields~\cite{synth1,dalibard}.

\paragraphtitle{Model parameters}
\Eq{e1} is formally like the Hamiltonian of coupled oscillators.  
Since the total number of particles is conserved, we have effectively ${d=M-1}$ degrees of freedom. 
The classical version of the BHH is obtained by defining occupation and conjugate phase
variables via ${\bm{a}_j = \sqrt{\bm{n}_j} \eexp{i\bm{\varphi}_j}}$. 
It is customary to define a scaled action-variable $\bm{I}_j=\bm{n}_j/N$.  
The classical equations of motion are known as the discrete non-linear Schrodinger equation (DNLS). 
In the continuous ${M \rightarrow \infty}$ limit they are known as the Gross-Pitaevskii equation.
The BHH can be described using the dimensionless parameters 
\be{2}
u = \frac{NU}{K}, \ \ \ \ \ \ \ \ \ \ \ \hbar=\frac{1}{N} 
\eeq
It should be recognized that $u$ is the ``classical" parameter of the model:
it is the only dimensionless parameter, except $M$ and $\Phi$, 
that appears in the DNLS equation.

\section{Metastability}

%
The semiclassical treatment, unlike the classical ``mean field" treatment, 
is in principle {\em exact}, provided the finite $\hbar$ is taken correctly into account.
Our main interest is in studying metastability of a prepared flow-state.
A~coherent flow-state is formed by condensing $N$ bosons into a single momentum orbital, 
and it is represented in phase-space by a Gaussian-like distribution of width $\hbar=1/N$, see \App{A}. 
Such state cannot maintain its phase-coherence if $\hbar$ is too large. Namely,
if the condition ${Mu \ll N^2}$ is not 
satisfied, one should expect a superfluid-Mott insulator transition, see e.g. \cite{sfr}.
Naturally we would like to avoid this regime. 
Still, it should be clear that the condition ${Mu \ll N^2}$ 
by itself is not sufficient in order to guarantee metastability 
of a prepared flow-state. 
The actual criterion for metastability requires a more serious semiclassical 
analysis, where ${\hbar}$ still plays an important role.
 
\begin{figure*}
\begin{center}
\begin{overpic}[width=0.45\hsize,tics=10]{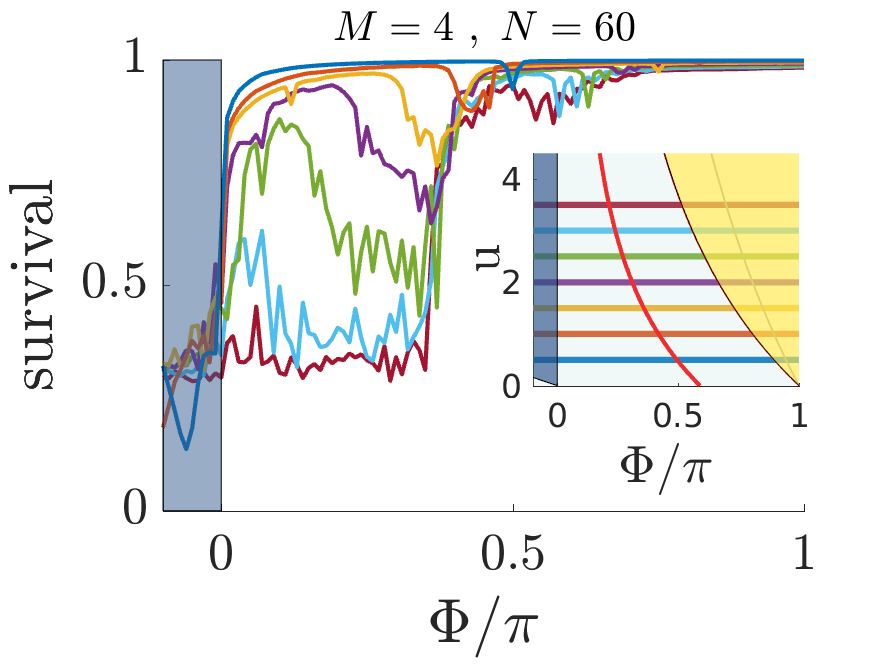}
 \put (19,71) {(a)}
\end{overpic}
\begin{overpic}[width=0.45\hsize,tics=10]{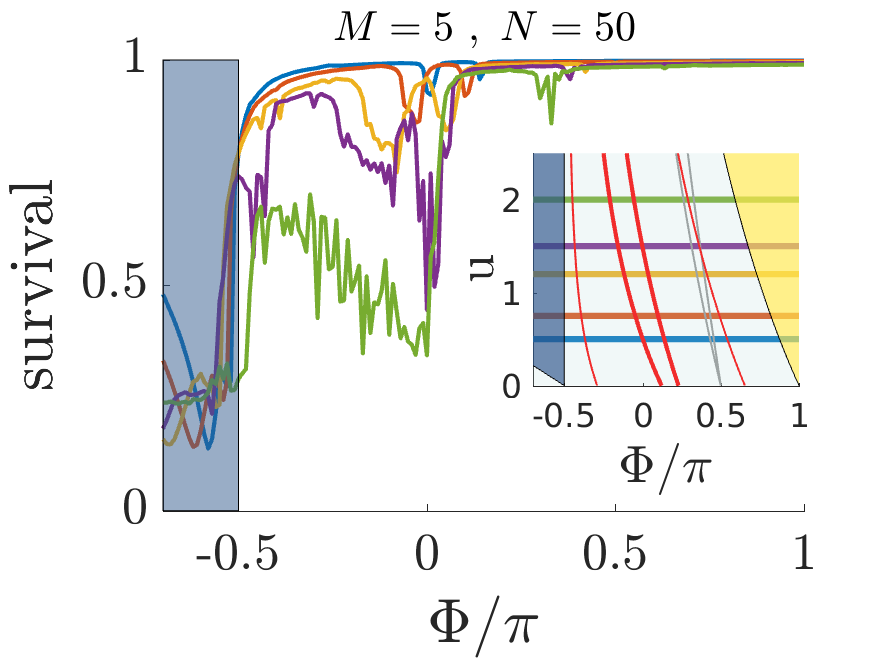}
 \put (19,71) {(b)}
\end{overpic}
\caption{ \label{fg2} 
{\bf The survival of a prepared coherent flow-state}. 
Quantum simulations are carried out in the region of interest that has been indicted by a green square in \Fig{fg1}.   
In each run all the particles are condensed initially in the same momentum orbital.  
The ``survival", see text, is plotted as a function of $\Phi$ for different $u$ values. 
The constant $u$ slices are indicated in the insets by the color-coded horizontal lines. 
The systematic shift of the dips and their broadening reflect the $u$ dependence of the non-linear resonances. 
Panels (a) and (b) are for 4-site and 5-site rings respectively.  
} 
\end{center}
\end{figure*}

\paragraphtitle{Linear stability analysis}
The stationary orbitals of a single particle in a clean ring are the momentum states, 
with winding number $m$ and wavenumber
\beq
k \ \ = \ \ \frac{2\pi}{M}m
\eeq
We can define occupation operators $n_k$, 
and write the BHH \Eq{e1} 
with $\bm{b}_k^{\dag}$ and $\bm{b}_k$ 
that create and destroy bosons in these momentum orbitals:
\be{3} 
\mathcal{H} \ = \ \sum_{k} \epsilon_k \bm{b}_k^{\dag}\bm{b}_k \ + \ \frac{U}{2M} \sum_{\langle k_1..k_4 \rangle} \bm{b}_{k_4}^{\dag}\bm{b}_{k_3}^{\dag}\bm{b}_{k_2}\bm{b}_{k_1} \ \ 
\eeq
Here $\epsilon_k = -K\cos(k-(\Phi/M))$, and the summation 
is over all the $k$ values that satisfy ${k_1+k_2=k_3+k_4}$ mod$(M)$. 
Condensing $N$ particles into the $m$th momentum orbital, 
we get the coherent flow-state 
\beq 
| m \rangle \ \ \equiv \ \ \frac{1}{\sqrt{N!}}\left( \bm{b}_{k_m}^{\dagger}  \right)^N | 0 \rangle 
\eeq
In the vicinity of this flow state one can perform a Bogoliubov approximation,   
see \App{B}. Consequently the quadratic part of the Hamiltonian  
takes the form 
\be{BT}
\mathcal{H}_0  \ = \ \sum_q \omega_q \bm{c}_{q}^{\dag} \bm{c}_{q} 
\ \equiv \ \sum_q \omega_q \tilde{n}_q 
\eeq
where $\tilde{n}_q$ are the occupation operators of the Bogoliubov quasi-particles, 
with creation operators $\bm{c}_q^{\dagger}$, such that ${\bm{b}_q^{\dagger} = u_q \bm{c}_q^{\dagger} + v_q \bm{c}_{-q}}$,  
where 
\be{4}
q=\frac{2\pi}{M} \ell, 
\ \ \ \ell=\text{integer}\neq0,  
\ \ \ -\frac{M}{2} < \ell \leq \frac{M}{2} 
\ \ \ 
\eeq
%
%
%
%
%
The coefficients of the transformation $u_q,v_q$ and the Bogoliubov frequencies $\omega_q$ are expressed as function of the model parameters ${(M,\Phi,u)}$, see \App{B} for details.

In \Fig{fg1} we present the implied stability diagram of the flow states. 
The horizontal axis is the unfolded phase
\beq
\phi \ \ =  \ \ \Phi - 2\pi m
\eeq
which allows us to address the stability of all the flow states in one extended diagram. 
Note that $\phi$ corresponds to the Bloch quasi-momentum $\phi/M$ 
in the analogous optical lattice problem \cite{niu}. 
The yellow color indicates Landau stability regions where all the Bogoliubov frequencies are {\em real} and have the same sign, 
implying that the flow state reside in an energetically stable island. 
The gray color indicates linear instability regions where some of the $\omega_q$ acquire an imaginary part, 
implying instability. We are left with an intermediate region that is conventionally regarded 
as dynamically stable \cite{niu}. But this linear-stability is in fact {\em fragile}  
and endangered by the non-linear terms of the Hamiltonian.

\section{The quench scenario}

We turn to test numerically whether a prepared coherent flow-state is metastable or not.
In a sense we simulate an hysteresis-type experiment, similar to that of \cite{NIST2}.
In the lab, the experimental protocol for a single run goes as follow: 
The first step sets the initial winding number 
using rotation frequency that corresponds to a Sagnac phase~$\Phi_0$, 
such that a Landau-stable flow state is prepared;
Then an instantaneous change of the rotation frequency 
is performed, and the Sagnac phase becomes~$\Phi$; 
After a waiting time~$t$ the BEC is released from the trap and imaged, 
consequently the final winding number is measured.

The quantum simulation of such quench scenario for the BHH ring 
is generated using a very efficient Leja interpolation procedure \cite{expleja}
that allows us to handle Hilbert space dimension up to~$\sim 10^5$. 
In the Landau stability regions we indeed observe perfect metastability, 
while in the linear unstable regime we observe exponential escape dynamics. 
In the later case, the momentum-orbital, where the bosons are initially condensed,   
is gradually depleted; while the occupation of the initially empty orbitals grows as $\exp(t/\tau)$, 
where $\tau$ is a constant that is related to the imaginary part of the Bogoliubov frequencies.       

We define the ``survival" as the normalized occupation of the condensate, 
as deduced from inspecting the long time dependence.
In \Fig{fg2} we show how the ``survival" of the condensate depends on~$\Phi$ 
for representative values of~$u$. Focusing on the intermediate regime, 
where the Bogoliubov analysis implies dynamical stability, 
we make the following observations:
{\bf (1)}~The dependence of the stability on~$\Phi$ is non monotonic, 
meaning that one can resolve dips in this dependence if~$u$ is small.
{\bf (2)}~Even in the center of a dip the stability is better than 
what could be expected when compared with the linear unstable regime.   
{\bf (3)}~These dips broaden and merge as~$u$ becomes larger.
Below we would like to provide an analytical explanation for the above observations.

\begin{figure*}
\begin{center}
\begin{overpic}[width=0.35\hsize,tics=10]{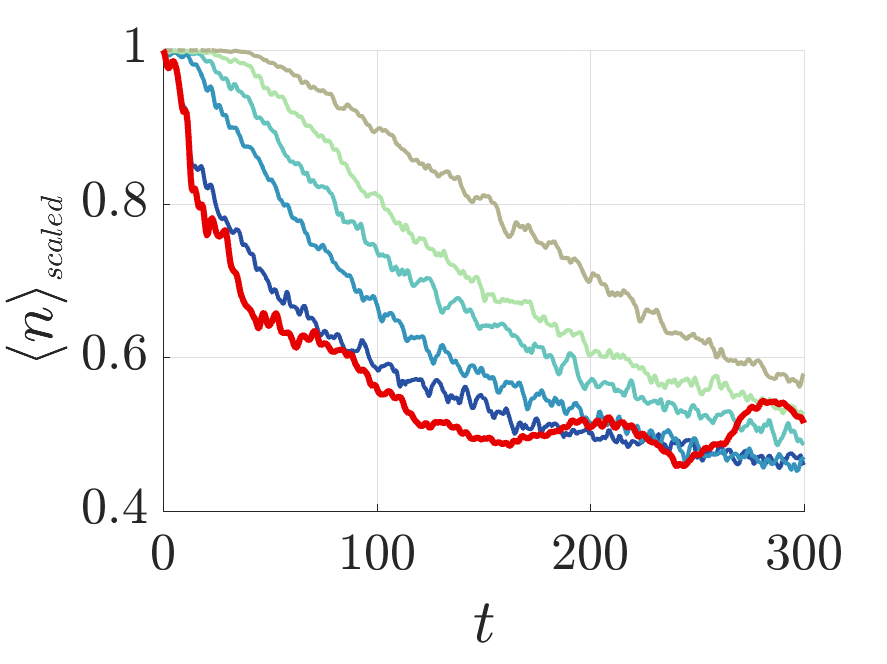}
 \put (50,72) {(a)}
\end{overpic}
\begin{overpic}[width=0.35\hsize,tics=10]{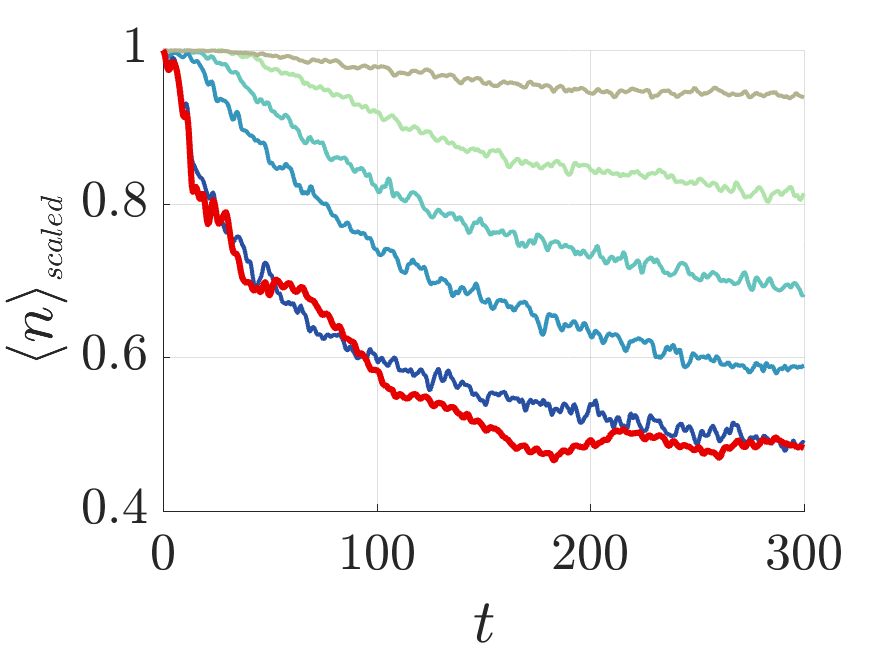}
 \put (50,72) {(b)}
\end{overpic}
\begin{overpic}[trim = 0mm -10mm 0mm 0mm, clip, width=0.21\hsize]{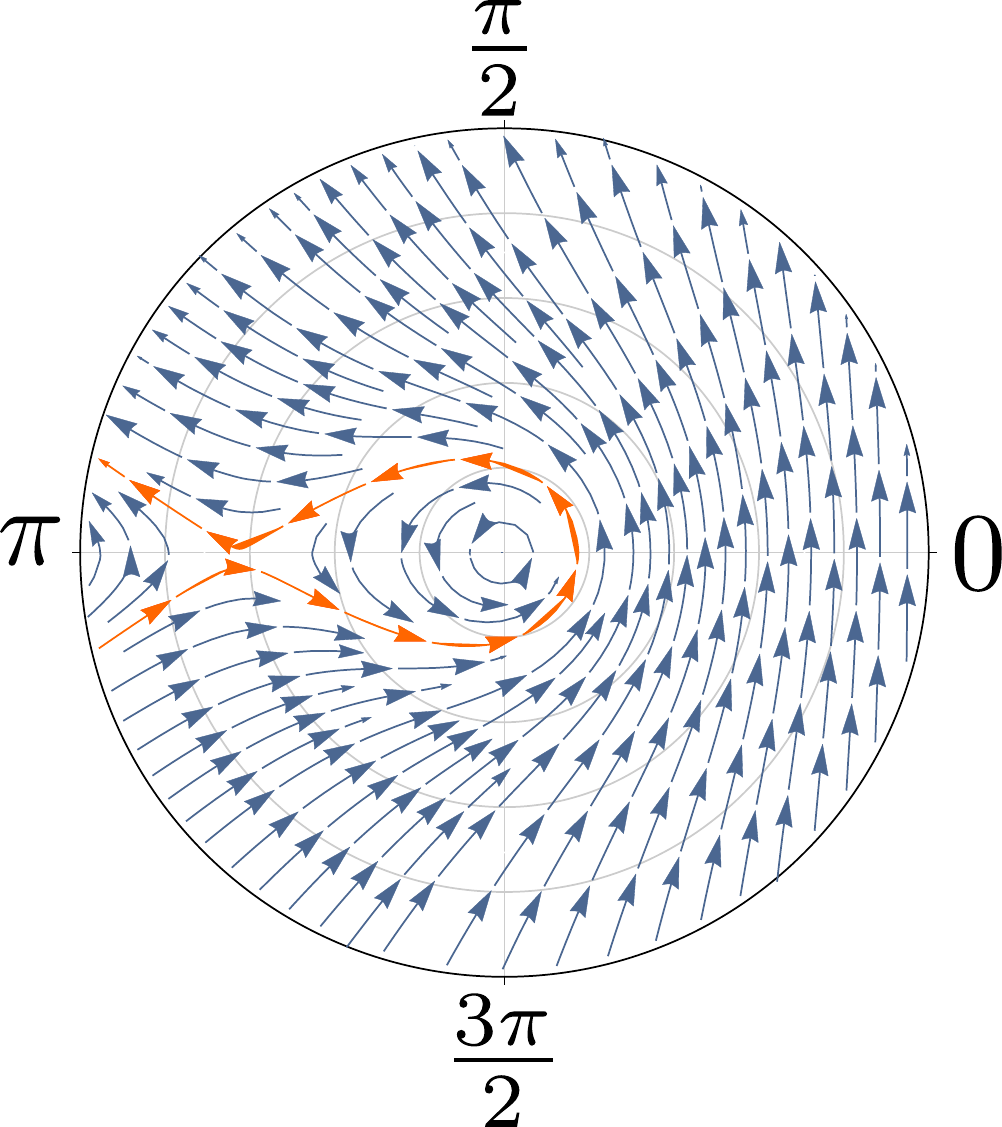}
 \put (10,85) {(c)}
\end{overpic}
\caption{ \label{fg3} 
{\bf The $N$ dependence of the decay process}.
Initially all the particles are condensed in the same momentum orbital.  
Panels~(a) and~(b) display the normalized occupation of the condensate as a function of time.
The red lines are quantum simulation for ${N=120}$ particles, 
while the other lines (blue to gray) are based on semiclassical 
simulations for ${N=120,500,1000,2000,4000}$ particles. 
The interaction strength is ${u=2.5}$. 
Panel~(a) relates to ${\Phi \approx 0.27 \pi}$, for which the detuning is ${\nu=0}$, 
while panel~(b) relates to ${\Phi =0.25 \pi}$. 
In the latter case we observe that the ``survival" becomes $N$ dependent.  
This reflects the existence of a stability island,  
as illustrated in panel~(c), which is a phase-space portrait of \Eq{e11}: 
the radial coordinate is $I\in[0,2.5]$; the polar angle is $\varphi$;  
and the parameters are ${\nu=2}$ and ${\mu=1}$ and ${J=1}$. 
} 
\end{center}
\end{figure*}

\begin{figure*}
\begin{center}
\begin{overpic}[width=0.4\hsize,tics=10]{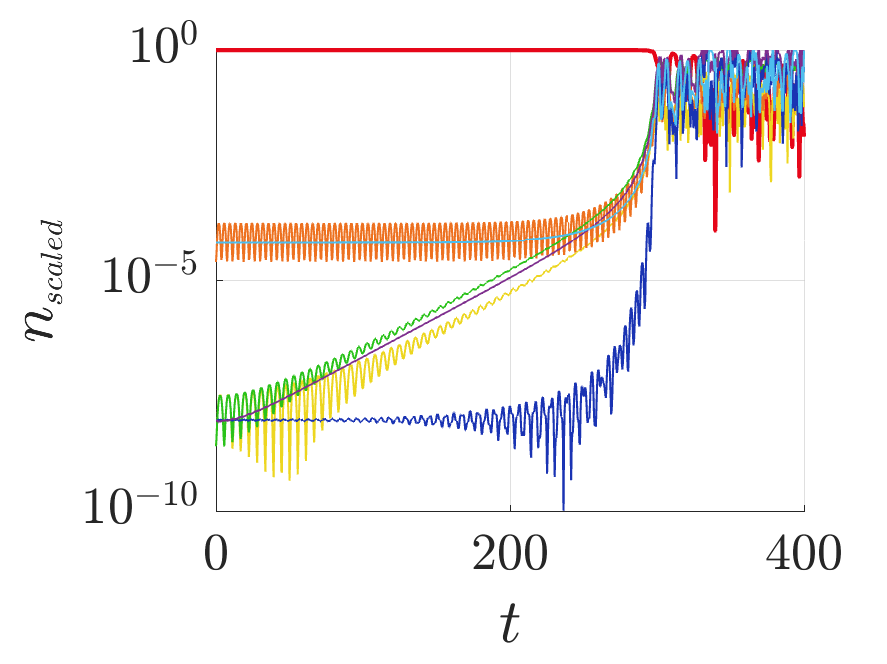}
 \put (25,72) {(a)}
\end{overpic} 
\begin{overpic}[width=0.4\hsize,tics=10]{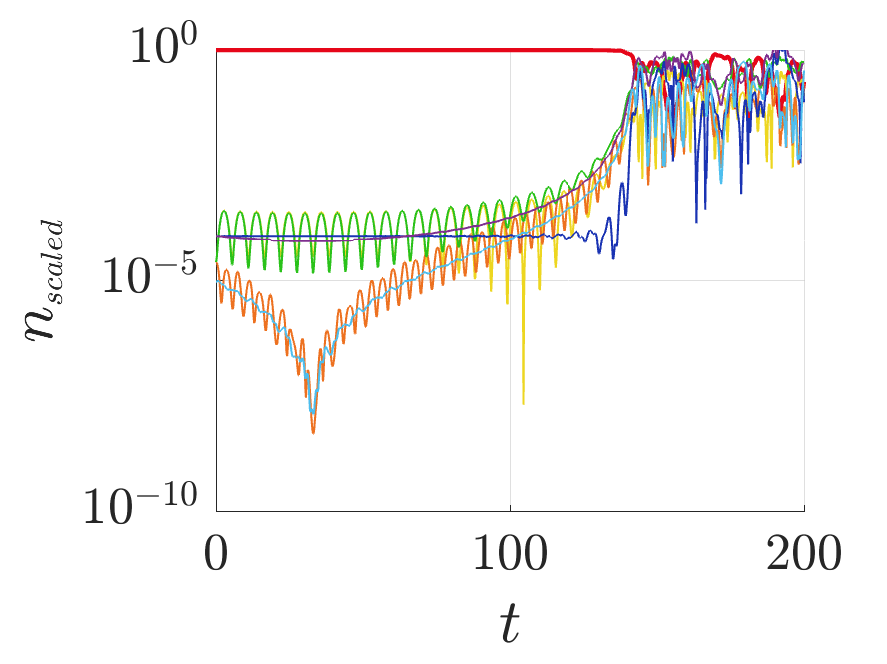}
 \put (25,72) {(b)}
\end{overpic} \\
\begin{overpic}[width=0.4\hsize,tics=10]{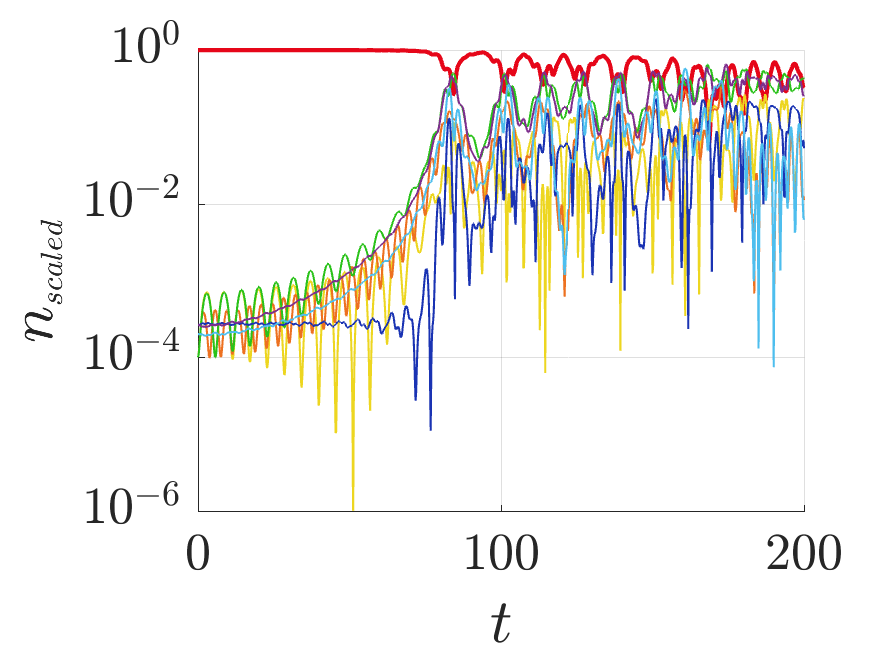}
 \put (25,72) {(c)}
\end{overpic}
\begin{overpic}[width=0.4\hsize,tics=10]{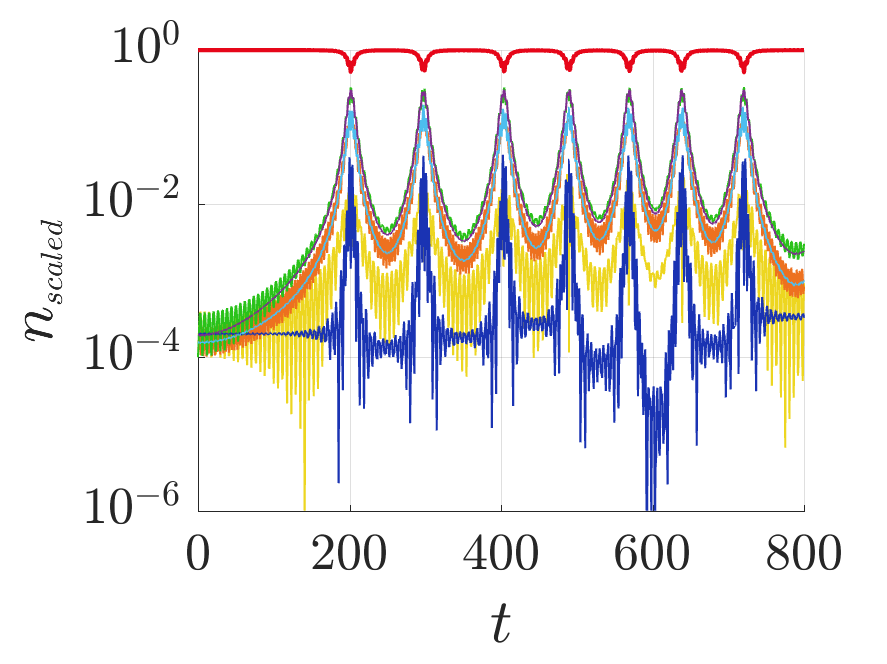}
 \put (25,72) {(d)}
\end{overpic}\\
\begin{Large}
\newcommand{\drawln}[3]{\begin{picture}(#2,1)(0,-3)\color[rgb]{#3}{\linethickness{#1mm}\line(1,0){#2}}\end{picture}}
\drawln{0.5}{22}{0.9000,0.0250,0.0980} $n_0$ ,
\drawln{0.5}{22}{0.9294,0.8392,0.1255} $n_1$ ,
\drawln{0.5}{22}{0.9294,0.4431,0.1255} $n_2$ ,
\drawln{0.5}{22}{0.1647,0.7608,0.1020} $n_3$ ,
\drawln{0.5}{22}{0.1000,0.2000,0.7000} $\tilde{n}_{1}$ ,
\drawln{0.5}{22}{0.4940,0.1840,0.5560} $\tilde{n}_{-1}$ ,
\drawln{0.5}{22}{0.3010,0.7450,0.9330} $\tilde{n}_{2}$
\end{Large}

\caption{ \label{fg4} 
{\bf The escape dynamics for zero detuning}.
Using classical simulation we plot the time dependence of the
momentum-orbital occupation $n_k$  (red, orange, green, yellow) 
as well as the quasi-particle occupations $\tilde{n}_q$ (blue, magenta, cyan) 
for a single trajectory that starts at ${n_0 \sim 1}$, 
while $n_k \ll 1$ for ${k\neq0}$. 
Note that the initial semiclassical cloud is composed of many such starting-points  
that occupy a Planck cell of radius $\sim 1/N$ in phase-space.
The dynamical scenario may consists of several steps, 
depending on the initial conditions. 
Typically it will start with either exponential stage~(a) or parabolic stage~(b).
Then it will cross to the hyperbolic escape of \Eq{e13} that terminates at $t_e$, 
possibly followed by chaotic motion~(a,b), 
or else there might be either intermediate or recurring re-injection scenario~(c,d).
The simulations are for an $M{=}4$ ring.  
In panels~(a,b) the interaction is ${u=3.84}$,  
while in panels (c,d) it is ${u=2.52, 1.58}$. 
The respective values ${\Phi=0.2\pi,0.27\pi,0.35\pi}$ 
have been adjusted via the resonance condition \Eq{e9}.     
} 
\end{center}
\end{figure*}
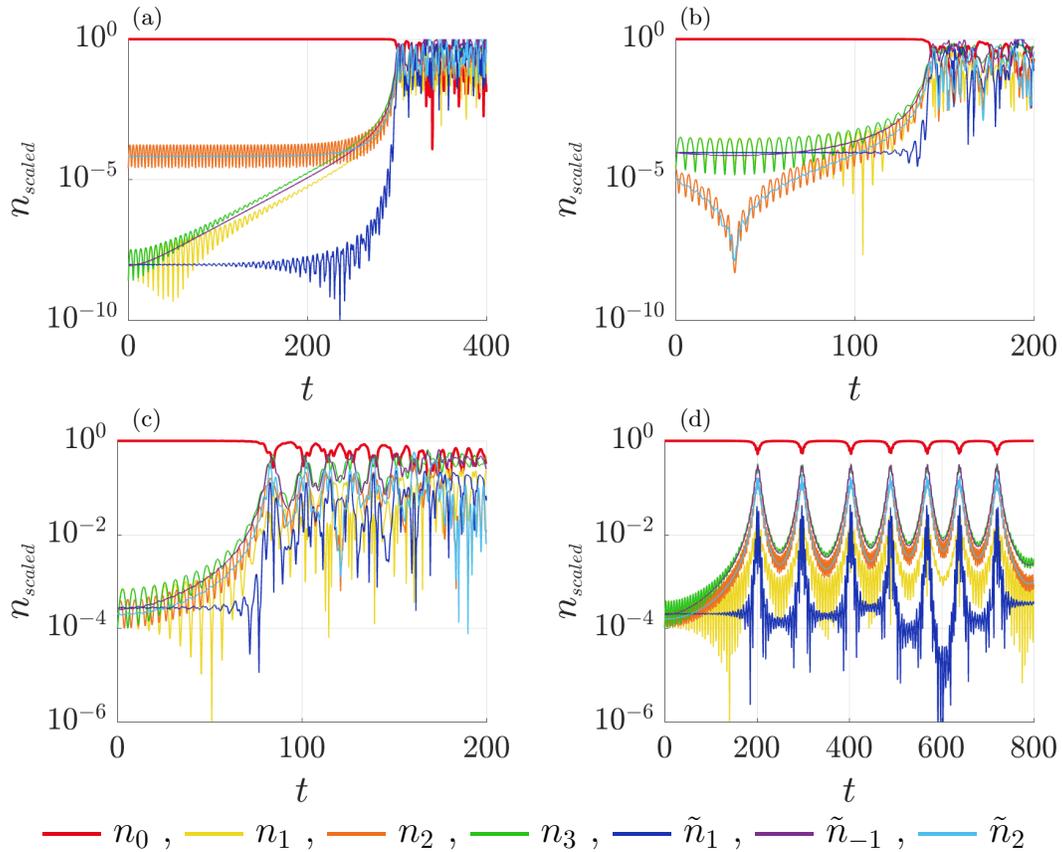

\section{Nonlinear resonances}

Including non-linear terms, the BHH
after the Bogoliubov transformation takes the form 
\be{6}
\mathcal{H}\ = \ \sum_q \omega_q \bm{c}_{q}^{\dag} \bm{c}_{q} +
\frac{\sqrt{N}U}{M}  \sum_{\langle q_1,q_2 \rangle} \left[V_{q_1,q_2}^{A} + V_{q_1,q_2}^{B}\right]
\eeq
where the summation is without repeating permutations. 
Above we have omitted the 4th order terms, see \App{C}, 
and kept only the 3rd order terms: 
\be{7}
V_{q_1,q_2}^{A} \ &=& \ A_{q_1,q_2} \left[ \bm{c}_{-q_1-q_2} \bm{c}_{q_2} \bm{c}_{q_1}  + \text{h.c.} \right] 
\\ \label{e8}
V_{q_1,q_2}^{B} \ &=& \ B_{q_1,q_2} \left[ \bm{c}_{q_1+q_2}^{\dag} \bm{c}_{q_2} \bm{c}_{q_1}  + \text{h.c.}\right]
\eeq
The coefficients $A$ and $B$ are functions of $(u,\Phi,M)$, 
and can be calculated explicitly, see \App{C}. 
The terms in $V^{B}$ are the so-called Beliaev and Landau damping 
terms \cite{DavidsonBogoliubov,Katz2002,Iigaya2006}, see also \cite{kolovskiPRE}, 
while the $V^{A}$ terms are usually ignored.
These terms can create resonance between the Bogoliubov frequencies. 
The resonance conditions are
\be{RC}
\omega_{q_1}+\omega_{q_2}+\omega_{-q_1-q_2}=0  \ \ \ \ & \mbox{for $V^{A}$} \\
\omega_{q_1}+\omega_{q_2}-\omega_{q_1+q_2}=0 \ \ \ \ \ & \mbox{for $V^{B}$} 
\eeq
We indicated these resonances in the diagrams of \Fig{fg1} 
using red and gray lines respectively. We realize in \Fig{fg2}
that the the $V^{A}$ resonances provide an explanation for the position of the dips:  
as $u$ is increased, their center is shifted in agreement with the diagram of \Fig{fg1}. 
Below we shall use action-angle variable representation 
to explain the effect of the resonances, see \App{D}.

\paragraphtitle{Proliferation of resonances} 
The $V^{A}$ terms do not endanger a Landau-stable flow state: 
in the energetically stable regime all the frequencies $\omega_{q}$ 
have the same sign, so the resonance condition of the $V^{A}$ terms 
cannot be satisfied (no red lines in the yellow regions of \Fig{fg1}). 
As for the the $V^{B}$ terms, they are important, if one is interested  
in the decay of a cloud of excitations that is released above 
a Landau-stable ground state \cite{DavidsonBogoliubov}.
{\em Contrary to that}, we establish below that the stability 
of the flow states outside of the Landau stability regime  
is mainly endangered by the $V^{A}$ terms, 
while in \App{E} we prove that the $V^{B}$ terms   
are not effective there.
 
For the $M{=}4$ device we realize that 
there is a single (red) resonance that is dominating, see \Fig{fg1}a. 
It is the ``1:2" resonance which involves ${q_1=q_2=q}$, 
such that \Eq{eRC} becomes ${2\omega_{q}+\omega_{-2q}=0}$.
For the ${m=1}$  flow state we set ${q=2\pi/4}$ and use \Eq{e5}, leading to:
\be{9}
u \ = \ 4 \cot \left(\frac{\Phi}{4} \right) \left[3\cos \left(\frac{\Phi}{4}\right)- \sqrt{6+2\cos \left(\frac{\Phi}{2}\right)} \right] \ \ \ 
\eeq
For larger rings more resonances appear, see \Fig{fg1}bc.  
In the $M\to\infty$ limit these resonances form as dense set, 
but are apparently of measure zero, because we know that 
that the non-discrete  ring limit is integrable \cite{ueda}.

\section{Secular approximation}

Keeping only the resonant term in the Hamiltonian, 
we can analyze, for example, the effect of the 
dominating ``1:2" resonance of \Eq{e9}  
(similar analysis applies to any resonance in an $M{>}4$ ring). 
This resonance couples the~$\omega_q$ mode
with the $\omega_{-2q} \equiv 2\omega$ mode. 
We define the detuning as ${\nu \equiv 2\omega_q+\omega_{-2q}}$.  
In action-angle variables, see \App{D},  
we end up with the Hamiltonian  
\be{11}
H_{q}  \ \ = \ \  \omega J  + \nu I + \mu I \sqrt{(J/2)+I}  \ \cos(\varphi)
\eeq
where $I$ is the normalized quasi-particle occupation $\tilde{n}_q$, 
and $\varphi$ is the conjugate phase, while $J$ is a constant of motion.
The interaction parameter is ${\mu=4|(NU/M)A|}$. 
The ${\nu=0}$ version of \Eq{e11} is the so called Cherry Hamiltonian of celestial mechanics \cite{Cherry}.

\paragraphtitle{The stability island}
Inspecting the phase-space of the Hamiltonian \Eq{e11} for a given~$J$ 
we realize that the $I$ motion near the origin is bounded 
provided the detuning  $|\nu|$ is large compared with ${\mu \sqrt{|J|}}$, see \App{F}. 
This is demonstrated in \Fig{fg3}c. 
From that we deduce that there is a stability island ${I,J < R_S}$ 
of radius ${R_S \sim (\nu/\mu)^2}$.
A coherent flow-state is represented in phase-space 
by a Gaussian-like distribution of radius $1/N$
that is centered around ${I=J=0}$.   
Accordingly such preparation becomes stable if $N$ is large enough, otherwise it decays. 
Consequently one deduces that the width of the instability region  
is determined by the condition 
\be{12}
\Big|\nu \Big| \ \ < \ \ A \left(\frac{1}{N}\right)^{1/2} \frac{u}{M}K
\eeq
This condition, via \Eq{e5}, 
provides a finite width to the resonance lines of \Fig{fg2}. 

The existence of a stability island should be reflected not only 
in the $N$ dependence  of the stability diagram, 
but also in the quench scenario itself.    
If the model parameters are tuned such that ${\nu=0}$  
we expect no stability, hence the ``survival" of particles in the condensate
should become {\em independent} of~$N$.  But for finite detuning, the survival 
should become larger if $N$ is larger: for smaller $\hbar$ a larger 
fraction of the Gaussian cloud is contained within the stability island.   
Using a quantum simulation it is not possible to confirm this prediction, 
because present day computing-power cannot handle much more than 100 particles, 
due to the exponential $N$ dependence  of the Hilbert space dimention.
Nevertheless, we are able to reconstruct the quantum simulations using a 
semiclassical procedure. Namely we propagate a cloud of trajectories in phase space. 
Larger $N$ means smaller cloud. In \Fig{fg3}ab we establish 
that the existence of a stability island can be detected via a quench experiment.
In such experiment the final occupation $\braket{n_k}$ of the flow-state momentum-orbital 
can be determined by a time of flight measurement, where ${t}$ is the waiting 
time after the quench. The $N$ dependence can be used 
as a measure for the size of a stability island.

\section{Decay and its suppression}

We can analyze the time dependence of the $\tilde{n}_q$ for trajectories 
that start outside of the stability island. 
The dynamical scenario may consists of several steps, 
depending on the initial conditions. 
Typically it will start with either 
exponential ($\tilde{n}_q \propto \exp(t/\tau)$), 
or parabolic ($\tilde{n}_q \propto (t-t_0)^2$) time dependence.
Then it crosses to hyperbolic escape:
\be{13}
\tilde{n}_q \ \ \propto \ \ \frac{1}{(t_e-t)^{2}}   \ \ \ \mbox{for $t<t_e$}
\eeq 
The term {\em hyperbolic growth} means that the action variable~$I$ of \Eq{e11}
grows towards a singularity within a finite time~$t_e$
that is determined by the initial conditions.
With the exact Hamiltonian the quasi particle occupations is of course bounded, 
hence the escape is followed by a chaotic motion, or else 
there might be a re-injection scenario. 
All these possibilities are illustrated in \Fig{fg4}, 
where we plot the time dependence of the momentum occupations~$n_k$,  
as well as the quasi-particle occupations $\tilde{n}_q$, 
based on exact DNLS simulations.

\paragraphtitle{Residual metastability}
We still have to explain the residual metastability that has been observed in \Fig{fg2}. 
Namely, why the ${\nu=0}$ dips of the ``survival" plot do not reach the same ``floor" 
that is implied by the simulations in the linear unstable regime.  
The answer pops into the eyes once we look in the lower panels of \Fig{fg4}. 
In these simulations the parameters were tuned such that ${\nu=0}$, 
so there is no stability island. Still for moderate values of~$u$ 
the cloud does not escape completely. The trapping here 
is due to the stickiness of phase space: many re-injections
are likely to happen before the trajectory escapes into the chaotic sea. 
The re-injection is due to remnants of integrable structures, namely,  
the Kolmogorov-Arnold-Moser (KAM) tori \cite{LLbook}.

\paragraphtitle{Types of metastability}
So far we have highlighted the quench scenario perspective of metastability, 
meaning that we start with an initial {\em coherent} flow-state, and ask about its ``survival".
But there is an optional perspective that is possibly more illuminating. 
In principle we can diagonalize the BHH, as in \cite{sfc}, 
and look for non-ergodic eigenstates that carry a large current. 
By investigating the structure of the underlying phase-space 
for those eigenstates, one deduces the following semiclassical classification:      
{\bf (i)}~Coherent flow-states that are supported by local minima of the energy landscape;
{\bf (ii)}~Coherent flow-states that are supported by quasi-integrable islands;
{\bf (iii)}~Chaotic flow-states that are supported by chaotic ponds in phase space;
{\bf (iv)}~Dynamically localized flow-states that are supported by sticky regions in phase space.
This classification is strictly well-defined for an $M{=}3$ ring, as discussed in \cite{sfc}, 
because such ${d=2}$ degree of freedom system has a mixed phase-space, 
where different regions (sea, islands, ponds) are separated by KAM tori. 
But in the present work we deal with high dimensional chaos (${d>2}$) 
for which the distinction between categories \mbox{ii-iii-iv} is blurred, 
because always in principle there is an escape option via 
Arnold diffusion \cite{LLbook,Basko2011,PhysRevLett.79.55,PhysRevLett.88.154101}
along the web of non-linear resonances. Still this is a very slow 
process and we have established that a secular approximation is 
capable of reproducing the essential physics of the superfluidity regime diagram.

\section{Discussion}

We have presented a semiclassical theory 
for the metastability regime-diagram 
of flow-states in BHH superfluid circuits; 
and for the decay of super-currents 
in hysteresis-type experiment that involve 
a quench scenario.  
We have explored how the metastability of the flow-state depends 
on the rotation frequency of the device, 
and on the strength of the inter-particle interaction.
\hide{
Depending on the dynamical regime one can distinguish between 
coherent, chaotic, and dynamically-localized flow-states.}
It is possible to use the same semiclassical framework 
to address the implications of introducing barriers 
or weak-links \cite{sfr}, which is complementary to the 
formal field-theory approach, see for example \cite{Minguzzi}.

The focus in the present work was on ${M>3}$ circuits 
that exhibit high-dimensional chaos \cite{ketzmerick1,ketzmerick2}.
This should be distinguished from the low-dimensional 
chaos of ${M=3}$ circuit, that has been analyzed in \cite{sfc}. 
The new ingredient in the analysis is the manifestation of non-linear resonances
that couple the Bogoliubov excitations. Contrary to the expectation they
do not originate from the Beliaev and Landau damping terms that create
and destroys pairs, but from the creation and destruction of triplets.
The former do not endanger the stability of the flow-states, 
because the dynamics under such perturbation remains bounded in phase-space. 
In contrast, the latter induce a decay process 
that is described by a variant of the Cherry Hamiltonian of celestial mechanics, 
and results in unbounded motion that leads to chaos and ergodization \cite{OlshErg}.

\appendix

\section{Semiclassical simulations}

The semiclassical simulations were done by propagating an initial cloud of $\sim$500 trajectories in phase-space.
Rescaling the field variables such that ${b_0=1}$, the initial coordiantes of the ${k\neq0}$ orbitals 
are written as ${b_k = r_k \eexp{\varphi_k}}$, where $r_k$ are picked randomly from a normal distribution 
with dispersion ${1/\sqrt{2N}}$, and $\varphi_i$ is a random phase.

\section{The Bogoliubov transformation}

Consider $N$ particles that are condensed into the same orbital~$m$.
In order to analyze the small excitations above this new ``vacuum",
the occupation of~$m$ is expressed as ${N-\sum_k \bm{b}_{k}^{\dagger}\bm{b}_{k}}$, 
where the sum excludes the ${k_m}$ orbital. 
Then the remaining unpaired operators $\bm{b}_{k_m}^{\dagger}$ and $\bm{b}_{k_m}$ 
are replaced by $\sqrt{N}$, and the quadratic part of the Hamiltonian  
is diagonalized into the form \Eq{eBT} by the Bogoliubov 
transformation ${\bm{b}_q^{\dagger} = u_q \bm{c}_q^{\dagger} + v_q \bm{c}_{-q}}$,  where
\be{10}
u_q, v_q =  \frac{1}{\sqrt{2}} \left(\frac{K_q+(NU/M)}{\sqrt{[K_q+2(NU/M)] \, K_q}} \pm 1 \right)^{1/2} 
\eeq   
The so-called Bogoliubov frequencies are:
\be{5}
\omega_{q} = K\sin(q) \sin\left(\frac{\phi}{M}\right) + \sqrt{\left(K_q+2\frac{NU}{M}\right)K_q} \ \ \ 
\eeq
where
\beq 
K_q \equiv 2K \sin^2\left(\frac{q}{2}\right) \cos\left(\frac{\phi}{M}\right)
\eeq
These frequencies are expressed as a function of the unfolded phase ${\phi = (\Phi - 2\pi m)}$.
Accordingly, without loss of generality we can assume ${m=0}$ and still address all 
the flow-states in one stability diagram.

\section{Non-linear terms}

We now wish to go beyond the quadratic terms, and include the non linear resonance terms which mix the Bogoliubov quasi particles.  Without loss of generality we assume condensation at ${m=0}$.  
Prior to the Bogoliubov transformation the unpaired operators $\bm{b}_0^{\dagger}$ and $\bm{b}_0$ 
were replaced by $\sqrt{N}$. The next term is obtained by picking the terms proportional to $\sqrt{N}$ in \Eq{e3}:
\beq
\mathcal{H}_{\text{res}} = \frac{\sqrt{N}U}{M} \sum_{\langle k_1..k_3 \rangle}
\left( \bm{b}_{k_3}^{\dag} \bm{b}_{k_2}\bm{b}_{k_1} + \bm{b}_{k_1}^{\dag}\bm{b}_{k_2}^{\dag}\bm{b}_{k_3} \right)
\eeq
where the summation 
is over all $k \neq 0$ values that satisfy ${k_1+k_2=k_3} $ mod$(M)$. 
Writing this expression in terms of the Bogoliubov quasi particle operators $\bm{c}_q^{\dagger}$ and $\bm{c}_q$ leads to \Eq{e6}.
The coefficients $A$ and $B$ are functions of $(u,\Phi,M)$, 
and can be calculated explicitly, see e.g. \cite{DavidsonBogoliubov}.
For ${A \equiv A_{q,q}}$ we get
\beq 
A = u_{2q}v_{q}^2+u_{q}^2v_{2q}+ 2u_{q}v_{q}(u_{2q}+v_{2q}) \ \ \ \ 
\eeq
There are also 4th order resonances that were neglected in \Eq{e6}.
Those are indicate in \Fig{fg1} by thin red lines. 
We did not consider them in the main text because they are 
suppressed by factor $\sqrt{N}$ compared with the 3rd order terms.

\section{Transforming to action-angle variables}

Keeping only the resonant terms in the Hamiltonian \Eq{e6}, 
we can analyze, for example, the effect of the ``1:2" resonance given by $V_{q,q}^{A}$.
Note that the treatment below holds for any~${M>3}$. 
The resonance couples the~$\omega_q$ mode
with the $\omega_{-2q} \equiv 2\omega$ mode. 
Under the secular approximation all the quasi-particle occupations 
except ${\tilde{n}_q \equiv \bm{c}_{q}^{\dag} \bm{c}_{q}}$ and $\tilde{n}_{-2q}$ 
are constants of motion and therefore ignored. Keeping only 
the $q$-related degrees of freedom we obtain:
\beq
\mathcal{H}_q = \omega_{q} \tilde{n}_q  +  2\omega \tilde{n}_{-2q} 
+  \frac{\sqrt{N}U}{M} A \left[  \bm{c}_{q}^2 \bm{c}_{-2q} + \text{h.c.} \right] \ \ \ \ 
\eeq 
In terms of action-angle variables ${\bm{c}_q = \sqrt{\tilde{n}_q} \eexp{i\varphi}_q}$
\beq
\mathcal{H}_q = \omega_{q} \tilde{n}_q  +  2\omega \tilde{n}_{-2q} + 2\frac{\sqrt{N}U}{M} A
\tilde{n}_{q} \sqrt{\tilde{n}_{-2q}}
\cos( \varphi ) \ \ \ \ 
\eeq 
where $ \varphi = 2\varphi_q + \varphi_{-2q}  $. Transforming to the 
variables ${I=\tilde{n}_q/(2N)}$ and ${J=(2\tilde{n}_{-2q}-\tilde{n}_q)/N}$, 
and scaling by $N$ the units of energy, we end up with \Eq{e11},  
where~$\varphi$ is conjugate to $I$, while $J$ is constant of motion.

\section{Ineffective resonances}

The $V^{B}$ terms are unable to destroy the stability of the flow state.
The proof is performed by applying a secular approximation, 
and transforming the interaction into action-angle variables, 
with $I=\tilde{n}_{q_1}$ and ${J=\tilde{n}_{q_1}+\tilde{n}_{q_2}}$. 
Observing that ${I \leq J}$, with $J$ being a constant of motion, 
it is implied that the action variable~$I$ cannot escape to ``infinity" 
and therefore stability is maintained.

\section{The stability island}

The action $I$ of \Eq{e11} is bounded from below by zero or by $|J/2|$  
depending on whether ${J}$ is positive or negative.
In the absence of detuning (${\nu=0}$) it will grow to infinity for all initial conditions. 
For a detuned system (finite $\nu$) a stability island can emerge, 
where the phase-space trajectories are bounded. 
The condition for its appearance is ${|\nu|>\mu\sqrt{(J/2)}}$ 
for ${J>0}$, and ${|\nu|>\mu\sqrt{(-3J/2)}}$ for ${J<0}$.   
The stability island is confined by a separatrix, see \Fig{fg3}, 
where the constant ${H_q(I,\varphi)}$ energy-surface possesses a saddle point. 
This saddle point is located along ${\varphi=\pi}$ for ${\nu>0}$, 
and along ${\varphi=0}$ for ${\nu<0}$, with
\beq
I_S = \frac{2}{9} \left[ \left(\frac{\nu}{\mu}\right)^2 - \frac{3 J}{2} + \sqrt{ \left(\frac{\nu}{\mu}\right)^4 +\frac{3 J}{2}  \left(\frac{\nu}{\mu}\right)^2 } \right]
\eeq
Accordingly the radius of the stability island, taking all possible~$J$ ``directions" into account, 
is ${R_S \sim (\nu/\mu)^2}$.

\clearpage



\end{document}